\documentclass[twocolumn]{article}

\begin{document}

\begin{titlepage}
\pagestyle{empty}
\bigskip
\title{\Large\bf Confinement and $U(1,3)$ symmetry of color particles\\
 in  complex phase space}
\medskip
\date{}
\bigskip
\smallskip
\author{\large V.V. Khruschev$^{a,b}$\thanks{e-mail: khru@imp.kiae.ru}\\
\textit{$^a$Russian Research Center '' Kurchatov Institute'', Kurchatov Square
1, }\\
\textit{ Moscow, 123182, Russia}\\
\textit{$^b$Department for gravitation researches in metrology, VNIIMS, Moscow,
Russia}\\}
\maketitle

\vspace*{1cm}

\noindent\textbf {Abstract} 
\vspace*{0.3cm}

It is shown that a universal confining potential 
for hadron constituents
can be  obtained  with the help of  $U(1,3)$ symmetry in 
a complex phase space.  Parameters of
this potential  are determined on the basis of spectroscopic data for
hadrons and results of lattice QCD calculations. We argue that the
account of the $U(1,3)$ symmetry  is important for
a description of strong interactions of quarks and gluons in 
a nonperturbative QCD domain at large interaction distances.

\medskip

\smallskip
\end{titlepage}
\setcounter{page}{2}

\noindent\textbf{1. Introduction}

\smallskip

 Color particles such as quarks and gluons are
observed in indirect measurements in the hadron physics, but they have never
been seen as usual particles in asymptotic free states. Although
perturbative QCD calculations of  processes at large momentum transfers
are confirmed by experimental data, rigorous calculations of many
nonperturbative effects are impossible at present even in the framework of
the lattice QCD (lQCD). The confinement of color particles is one amongst
the main phenomena of the nonperturbative QCD (nQCD) and a good deal of
effort are demanded for incorporating  it into a complete theory of strong
interactions \cite{swanson}. In spite of the fact that the confinement, perhaps, will
be validated with the QCD Lagrangian, this property of QCD can be connected
with some symmetry, which has not been taking into account yet. For
instance, it may be a kinematical symmetry connected with generalized
space-time properties of quarks and gluons.  It is well known that the
Poincar\'e symmetry, which is the space-time symmetry of the orthodox
relativistic quantum field theory, is originated from the isotropy and the
homogeneity of the Minkowski space-time and is based upon observations of
macro- and micro-phenomena concerning ordinary physical bodies and particles.

 In the present paper we suppose that the confinement of color
particles can be understood as a corollary of a generalized symmetry, which
possess nonperturbative interactions of color particles. We consider the $%
U(1,3)$ group as the group of symmetry of such kind. It is possible, that
the $U(1,3)$ symmetry belong to a class of approximated or exact nQCD
symmetries. In any case the $U(1,3)$ symmetry in a complex phase space
allows easily to include the confinement of color particles into
consideration and in the framework of potential models give an explicit form
of a confining potential.

 The $U(1,3)$ symmetry has been used in physics for the first time as
the group of a dynamic symmetry of  a isotropic oscillator \cite{mukunda}. Its
discrete unitary representations were applied for a description of hadron
properties \cite{kalman, saveliev}. In Ref. \cite{kunst} the $U(1,3)$ symmetry of a complexified
theory of gravitation has been found. This symmetry  has been proposed in
Refs. \cite{khrusch} as a extended symmetry in a complex phase space for quarks and
gluons. $SU(1,3)$ symmetrical coherent states were studied as well \cite{gitman} and a
generalized $U(1,3)$ symmetrical quantum mechanics  was considered in Refs.
\cite{low}.

 The  paper  is organized as follows: sections 2 and 3 are
devoted to a brief description of $U(1,3)$ transformations and a realization
of their representations on generalized quark fields. An $U(1,3)$
symmetrical confining potential and possible values of its parameters are
given in section 4. Conclusions are found in section 5.

\smallskip

\noindent{\bf 2. $U(1,3)$ symmetry}

\smallskip

 The pseudounitary $U(1,3)$ group can be defined as a group
of transformations in a complex phase space $C_{4}$ with vectors 
$c_{\mu}=q_{\mu}-i\kappa^{-1}p_{\mu}$, $\mu =0,1,2,3 $, which leave
invariant the following Hermitian form:
\begin{equation}
|c|^{2}=c_{\nu }c_{\mu }^{\ast }\eta ^{\mu \nu }=c_{0}c_{0}^{\ast
}-c_{1}c_{1}^{\ast }-c_{2}c_{2}^{\ast }-c_{3}c_{3}^{\ast },  \label{form1}
\end{equation}

\noindent where $\ast $ is a complex conjugation,  $\eta ^{\mu \nu
}=diag\{1,-1,-1,-1\}$, $\eta ^{\mu \nu }\eta _{\nu \sigma }=\delta _{\sigma
}^{\mu }$, $\kappa $ is a constant with dimensions of  $[M^{2}]$ in the
natural system of units with $\hbar =c=1$. It is convenient to define
contravariant vectors : $c^{\mu }=\eta ^{\mu \nu }c_{\nu }^{\ast }$ ,  then
the invariant Hermitian form (\ref{form1}) can be written as $c_{\mu }c^{\mu%
}$.    Translations in
the complex space $C_{4}$ are denoted through $m_{\mu}=p_{\mu}+i\kappa q_{\mu}$.

 Let us consider along with the complex phase space $C_{4}$
a  complex Grassman algebra $G_{4}$ with generating elements $\eta ^{\alpha%
}$, $\alpha =0,1,2,3$ and a set of functions $F_{i}$, which are defined
on $C_{4}$ $\otimes $ $G_{4}$ and have the following form:
$$
F_{i}(c^{\mu },c_{\mu };\eta ^{\alpha })=f_{i}(c^{\mu },c_{\mu
})+
$$
\begin{equation}
\sum_{k>0}\sum_{\alpha _{1}<...<\alpha _{k}}f_{i\alpha _{1}...\alpha
_{k}}(c^{\mu },c_{\mu })\times \eta ^{\alpha _{1}}...\eta ^{\alpha _{k}}
\label{graf}
\end{equation}


 One can choose a subset $F_{iD}$, consisting of those
functions $F_{i}$ , which satisfy an $U(1,3)$ invariant equation of the
Dirac type \cite{khrusch}:

\begin{equation}
(i\eta ^{\mu }\frac{\partial }{\partial c^{\mu }}+i\frac{\partial }{\partial
\eta ^{\mu }}\frac{\partial }{\partial c_{\mu }}-\frac{C}{2})F(c^{\mu
},c_{\mu };\eta ^{\alpha })=0  
\label{dirf}
\end{equation}


In the quantum case, when $ q_{\mu }$ and  $p_{\mu }$ are noncommutative
quantities, one should consider functions only in a $\ $coordinate
representation or in a momentum representation. Thus we perform a
substitution from  $c^{\mu }$ and $ c_{\mu }$ variables to $ q_{\mu }$
and  $p_{\mu }$ variables and restrict ourselves, for instance, only those
functions, which depend on coordinates and transform under irreducible
representation (IR) of the $U(1,3)$ group. The following relations are
useful, when one make this substitution:
$$
\frac{\partial}{\partial c^{\mu }}=\frac{1}{2}
(\frac{\partial}{\partial q^{\mu}}-i\kappa\frac{\partial}
{\partial p^{\mu}}),
\frac{\partial}{\partial c_{\mu }}=\frac{1}{2}(\frac{\partial}{\partial q_{\mu }}%
+i\kappa \frac{\partial }{\partial p_{\mu }}).  \label{derf}
$$
\begin{equation}
\gamma _{\mu }=\eta _{\mu }+\frac{\partial }{\partial \eta ^{\mu
}},\quad \gamma _{\mu }^{\prime }=i(\eta _{\mu }-\frac{\partial }{%
\partial \eta ^{\mu }}),  
\label{gamf}
\end{equation}


\noindent where \{$\gamma _{\mu},\gamma _{\nu }\}=2\eta _{\mu \nu }$, \{$\gamma _{\mu
},\gamma _{\nu }^{\prime }\}=0.$

\smallskip

\noindent{\bf 3.  Realization of the $U(1,3)$ symmetry on generalized
quark  fields}

\smallskip

 If we take into account that any quark
belong some hadron, thus quarks (or other color particles) should be
described with bilocal fields $\Psi _{i}(q_{\mu },Q_{\mu })$, where $q_{\mu
} $ are quark coordinates and $Q_{\mu }$ are hadron ones. In order to
satisfy the translation invariance  condition   the fields $\Psi
_{i}(q_{\mu },Q_{\mu })$ are \ reduced to fields $\Psi _{i}^{\prime }$,
which depend on differences $q_{\mu }-Q_{\mu }$. Analogously, \ an
interaction potential, which one can add in the equation (\ref{dirf}),
should be dependent on the $q_{\mu }-Q_{\mu }$. In the following we restrict
ourselves by a consideration of a four-dimensional fundamental $U(1,3)$
IR, which is a bispinor representation of the
Lorentz group as well. Then for a generalized quark field $\psi (q_{\mu
}-Q_{\mu })$, using (\ref{dirf}), (\ref{derf}) and (\ref{gamf}), one can write the Dirac equation in the following form:
$$
(i\gamma ^{\mu }\frac{\partial }{\partial (q-Q)^{\mu }}-
\gamma ^{\mu
}V_{\mu }(q_{\mu }-Q_{\mu })-
$$
\begin{equation}
S(q_{\mu }-Q_{\mu })-m)\psi (q_{\mu }-Q_{\mu
})=0,  \label{dirfin}
\end{equation}


\noindent where $V_{\mu }(q_{\mu }-Q_{\mu })$ is a Lorentz vector part of an
interaction potential, while $S(q_{\mu }-Q_{\mu })$ is a Lorentz scalar part.

\smallskip 

\noindent{\bf 4.  $U(1,3)$ symmetrical confining potential}

\smallskip

For a $U(1,3)$ invariant theory in $C_4$ a Poincar\'e mass for a quark $m$  is
coordinate dependent and should satisfied the following equation:
\begin{equation}
m_{C}^{2}=m^{2}(q_{\mu }-Q_{\mu})+\kappa^{2}(q_{\mu }-Q_{\mu})(q^{\mu
}-Q^{\mu })  
\label{newmss}
\end{equation}


Eq. (\ref{newmss}) allows to define the coordinate dependence of a quark mass and
gives the form of a confining potential unambiguously. Moreover, the
confining potential is a scalar with respect to the Lorentz transformations.
This fact is in accordance with allowable transformation properties of a
confining potential in the QCD \cite{grom}.

 For calculations of bound state characteristics one can use a
simultaneous approximation because of, as usual, steady characteristics of a
bound state are needed. If $N$ particles \ with coordinates $q_{i\mu }$, $%
i=1,...,N,\mu =0,1,2,3$ interact each other and $\ P$ is their total
momentum, then one can impose  constraints on a range of variables $q_{i\mu
} $: $P^{\mu }q_{1\mu }=P^{\mu }q_{2\mu }=...=P^{\mu }Q_{\mu }$. Thus in
the rest frame of a hadron the static $U(1,3)$ invariant \ potential for
each $i$-th quark is:
\begin{equation}
V_{Si}((\mathbf{q}_{i}-\mathbf{Q})^{2})=\sqrt{m_{Ci}^{2}+\kappa^{2}
(\mathbf{q}_{i}-\mathbf{Q})^{2}}  
\label{scapot}
\end{equation}

The scalar potential (\ref{scapot}) provides the confinement for color
particles, which is linearly growing at large distances $|\mathbf{q}_{i}-%
\mathbf{Q|}$ \ with the coefficient equal to the parameter $\kappa $. If
one consider only two interacting particles, for instance, a quark and an
antiquark, then the confining potential behave as $\kappa r$ at large
separations $r$ between a quark and an antiquark. In the usually \
accepted notation $\kappa =\sigma $, where $\sigma $ is the so called
string tension. According to the results of Ref. \cite{khru}, the numerical value
of $\sigma $ is equal to $0.20\pm 0.01$ $GeV^{2}$, however, if one include
results obtained in other  models \cite{simon}, then the error value
should be enlarged  $\sigma =$ $0.20\pm 0.02$ $GeV^{2}$ .  If $\sigma $
is known it is possible to find two constants concerning interactions in a
confinement domain, according to the following relations: $m_{c}l_{c}=1$, \ $%
\ \sigma =m_{c}/l_{c}$, i.e. $m_{c}$ and $l_{c}$, with dimensions of the
mass and the length, respectively. Taking into account the numerical value
of $\sigma $ written above one can obtain that \ the confinement mass $m_{c}$
is equal to $0.45\pm 0.02$ $GeV$, while the confinement length $\ l_{c}$
is equal to $0.44\pm 0.02$ $Fm$. \ It is consistent with the facts that
a hadron formation starts when a string length between opposite color
charges become larger or of the order of \ $l_{c}$ and typical transversal
momenta \ $p_{\perp }$ of \ hadrons produced are of the order of $m_{c}$.

 In the framework of potential models a vector part of an interaction
potential, for instance, between a quark and an antiquark, is taken in the
quasi-Coulombic form: $V(r)=-4\alpha _{s}/3r$, where $\alpha _{s}$ is a
strong interaction constant at small interaction distances. Thus, if we take
the same vector potential between a quark and an antiquark, and the sum of
two scalar potentials (\ref{scapot}) with $U(1,3)$ constants $\ m_{C1}$ for a quark
and $m_{C2}$ for a antiquark, then the total static potential can be written
as
$$
V_{tot}(r)=-4\alpha_{s}/3r+C_{V}+\beta (C_{S}+
$$
\begin{equation}
\sqrt{m_{C1}^{2}+(\kappa
r/2)^{2}}+\left.\sqrt{m_{C2}^{2}+(\kappa r/2)^{2}}\right),  
\label{finpot}
\end{equation}
\noindent where $\beta $ is the known Dirac matrix, $C_{V}$ \ and $\ C_{S}$ are two
arbitrary constants for vector and scalar parts of  $V_{tot}(r)$, $\ $%
correspondingly. Note, that the dependence of the scalar potential versus $r$
is not strictly linear, so if we expand $V_{S}(r)$ at large values of $r$,
the following expression can be obtained:
\begin{equation}
V_{S}(r)=C_{S}+\sigma r+\frac{m_{C1}^{2}+m_{C2}^{2}}{\sigma r}+...
\label{apscap}
\end{equation}
It is interesting to identify  the third term in the expansion (\ref{apscap}) 
with the L\"uscher term, which is verifiable by lQCD calculations \cite{lush}. 
In this case the sum $m_{C1}^{2}+m_{C2}^{2}$ is equal to $\ -\pi \sigma /12$.

\smallskip

\noindent\textbf{5. Conclusions}

\smallskip

 We conclude that the confinement of color particles is easily
incorporated into the framework of $U(1,3)$ invariant  models with the
square root scalar potentials (\ref{scapot}). The parameters of these
potentials can be estimated with the help of potential model, string model
and lattice QCD results. The asymptotic confining potential   is linearly
growing at large distances and is invariant under the Lorentz group and a
change of quark flavors. Moreover the expansion of the confining potential 
at large distances gives the term, which can be identified with the well
known L\"uscher term. These facts give the support for the importance of the
account of the $U(1,3)$ symmetry in the complex phase space $C_{4}$
for interactions of
quarks and gluons at large distances in the nonperturbative QCD domain.

\smallskip

\textbf{Acknowledgements}.\ \ 


The author is grateful to Yu.V.~Gaponov,  A.N.~Leznov, V.I.~Savrin, 
S.V.~Semenov and A.M.~Snigirev for useful discussions.



\begin{thebibliography}{99}
\bibitem{swanson}
 E.S. Swanson, Plenary talk at the Intern. Conf. \textit{\ }''Hadron
2003'', Aschaffenburg, Germany\textit{\ \ }[arXiv:hep-ph/0310089].
\bibitem{mukunda} N. Mukunda, L.O'Raifeartaigh, and E.C.G. Sudarshan, Phys. Rev. Lett. 15
(1965) \ 1041.
\bibitem{kalman} C.S. Kalman, \ Can. J. Phys. 51 (1973) 1573.
\bibitem{saveliev}
 M.V. Saveliev and V.V. Khruschev, Yad. Fiz. 22 (1975) 1253.
\bibitem{kunst}
 G. Kunstatter, R.Yates, J.Phys. A14(1981) 847.
\bibitem{khrusch}
 V.V. Khruschev, Yad. Fiz. 46 (1987) 219, ibid. 58 (1995) 1869. Preprint
IHEP 85-120, Serpukhov, 1985.
\bibitem{gitman}
 D. M. Gitman and A.L. Shelepin, J. Phys. A26 (1993) 7003.
\bibitem{low}
 S.G. Low, Nuovo Cim. B108 (1993) 841, \ J. Math. Phys. 38 (1997) 2197, J.
Phys. A35 (2002) 5711.
\bibitem{grom}
 D. Gromes, Phys. Lett. B202 (1988) 262.
\bibitem{khru}
 V.V. Khruschev, V.I. Savrin and S.V. Semenov, Phys. Lett. B525 (2002)
283 [arXiv:hep-ph/0111055].
\bibitem{simon}
 Yu.A. Simonov, Invited talk at the Intern. Conf. dedicated to the 90-th
birthday of the late Prof. I.Ya. Pomeranchuk [arXiv:hep-ph/0310031].
\bibitem{lush}
 M.L\"uscher and P. Weisz, JHEP 0207 (2002) 049 [arXiv:hep-lat/0207003].
\end{thebibliography}
\end{document}